\documentclass[12pt]{article}
\usepackage{amsmath}
\usepackage{amsfonts}
\usepackage{graphicx}
\usepackage{enumerate}
\usepackage{natbib}
\usepackage{url} 
\usepackage[colorlinks = true, linkcolor = red!50!black, citecolor = blue, urlcolor = green!50!black]{hyperref}
\usepackage{caption}
\usepackage{pdfpages}

\usepackage{longtable}
\usepackage{booktabs}

\usepackage{bm}
\usepackage{amsthm}
\usepackage{mathtools}
\usepackage{tikz}
\usetikzlibrary{calc, decorations.pathreplacing, arrows.meta}

\def\bSig\mathbf{\Sigma}

\newcommand{\blind}{1}

\begin{document}

\def\spacingset#1{\renewcommand{\baselinestretch}%
{#1}\small\normalsize} \spacingset{1}


\if1\blind
{
  \title{\bf PEP: a tackle value measuring the prevention of expected points}
  \author{Robert Bajons\thanks{e-mail: rbajons@wu.ac.at.},\hspace{.2cm}\\ 
    Jan-Ole Koslik, \\ 
    Rouven Michels, \\ 
    Marius Ötting
    }
  \maketitle
} \fi


\if0\blind
{
  \bigskip
  \bigskip
  \bigskip
  \begin{center}
    {\LARGE\bf }
\end{center}
  \medskip
} \fi

\bigskip
\begin{abstract}
\noindent
Traditional assessments of tackling in American Football often only consider the number of tackles made, without adequately accounting for their context and importance for the game. Aiming for improvement, we develop a metric that quantifies the value of a tackle in terms of the prevented expected points (PEP). Specifically, we compare the real end-of-play yard line of tackles with the predicted yard line given the hypothetical situation that the tackle had been missed. For this, we use high-resolution tracking data, that capture the position and velocity of players, and a random forest to account for uncertainty and multi-modality in yard-line prediction. Moreover, we acknowledge the difference in the importance of tackles by assigning an expected points value to each individual tree prediction of the random forest. Finally, to relate the value of tackles to a player's ability to tackle, we fit a suitable mixed-effect model to the PEP values. Our approach contributes to a deeper understanding of defensive performances in American football and offers valuable insights for coaches and analysts.
\end{abstract}

\noindent%
\vfill

\newpage
\spacingset{1.9} 

\section{Introduction}

%

The emergence of high-resolution tracking data has initiated a new era of sports analytics. Across numerous sports (for a detailed review, see \citealp{kovalchik2023player}), the positions, velocities, and accelerations of individual players enable enhanced player and game situation evaluation \citep{goes2021tactics, muller2021pivot}, clustering of game plays \citep{ChuReyersThomsonWu+2020+121+132} or decoding of tactics \citep{michels2023nonparametric,otting2023football}. 
Although academic researchers have primarily conducted these analyses, the rising number of employees within sports clubs dealing with data \citep{vanhaaren} and the distribution of tracking data (e.g., for the German Bundesliga, see \citealp{dfl}) underscore its emerging practical relevancy in daily decision-making processes.

However, a common property of the investigations of tracking data is the focus on the offense of teams, while defensive performances have received less attention \citep{forcher2022use}. 
This is especially problematic in sports such as American football, in which offense and defense are strictly separated, thus making an isolated analysis of the latter feasible. However, in American football, interest primarily was laid on decision-making of the attacking team \citep{heiny2011predicting,joash2020predicting, adam2024markov}
or offensive player and game play evaluation (see e.g.\ \citealp{deshpande2020expected,eager2023investigating,nguyen,reyers2021quarterback}). In contrast, there is limited literature on the analysis of defensive actions \citep{DuttaYurkoVentura+2020+143+161, nguyen2024fractional}. Inspired by the NFL Big Data Bowl 2024, we take a closer look at the defensive performance of players. Specifically, we aim to assign a value to every single tackle made by an individual player with the help of our new metric PEP (prevented expected points).

\begin{figure}
    \centering
    \includegraphics[width = 1\textwidth]{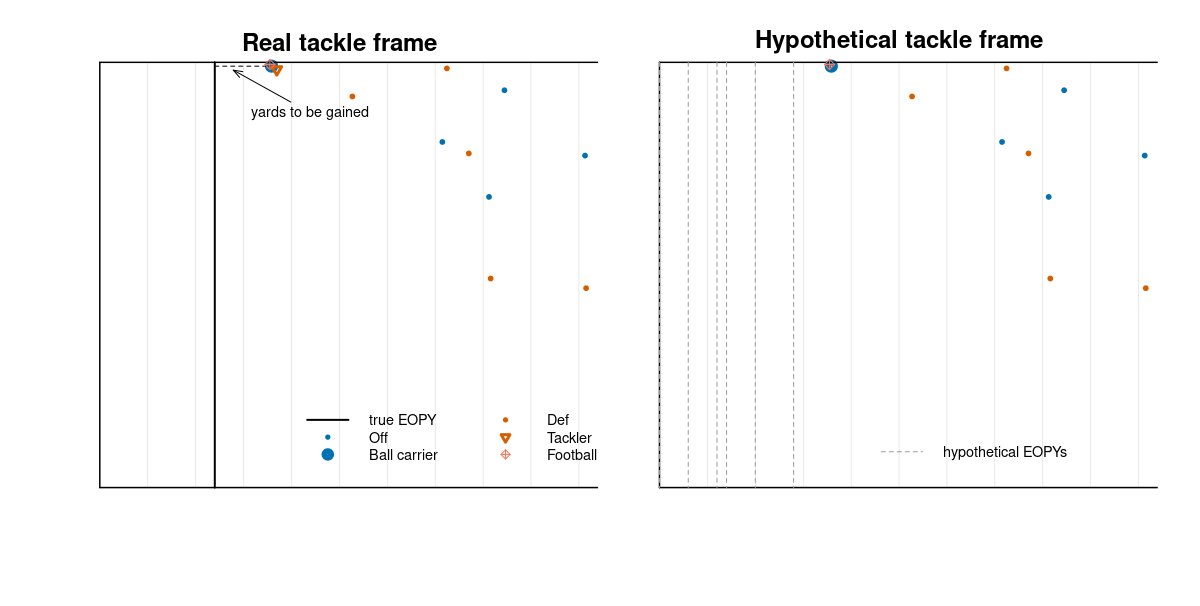}
    \caption{The left panel displays the real tackle frame from the data. We indicate the real EOPY after the defensive player tackled the offensive player. In contrast, the right panel shows the hypothetical tackle frame, in which the tackler is removed and the hypothetical EOPY must be estimated.}
    \label{fig:real}
\end{figure}

For defining this value, we construct the hypothetical scenario that a tackle --- which in reality took place --- was missed by a player and aim to quantify the yards saved by a defensive player. Ideally, albeit impractically, running a play twice --- once with the defense player executing the tackle and a second time without --- would allow a direct comparison of the yardage gained by the ball carrier, hence enabling evaluating of the defensive player’s tackle. 
However, given the impracticability of such a hypothetical scenario, we suggest an approach that involves approximating this scenario by predicting the yard line of the ongoing play if the closest defender (who executed the tackle) had missed the tackle. In terms of statistics that means excluding this player from the data at the moment of the tackle and predicting the end-of-play yard line (EOPY) in this hypothetical game situation (see Figure~\ref{fig:real} for an exemplary play). In fact, this simplification is a realistic reflection of \textit{real} missed tackles, as those are often characterized by a reduced speed of the offensive player, similar to what we observe when removing the nearest defender.

When predicting the EOPY, we aim to account for uncertainty and multi-modality in the prediction. Thus, we decided to produce a full conditional density estimate using a random forest instead of a sole mean prediction of the EOPY (see \citealt{yurko2024nfl} for a similar approach). However, only quantifying the yards saved by a particular tackle is not satisfactory as an adequate measure of tackle value. For example, consider two scenarios: 1) it is 4th down and the opponent is two yards away from the own end zone and 2) it is 1st down and the opponent is in its own half. Comparing the two situations, a tackle that saves two yards is much more valuable in the first scenario than a tackle that saves two yards in the second scenario. 
Therefore, we aim to produce a measure of tackle value on the scale of expected points (EP), which is a concept that has seen a steady rise in the last decade, among other things for its use for 4th down decision making \citep{yam2019lost, brill2023analytics} or player evaluation \citep{yurko2019}. EP can be viewed as a complex mapping of the EOPY to the points that are expected to be achieved from the next play onwards. 
Combining this idea with the conditional density estimation of the EOPY, we can calculate the mean 
expected points in any given game situation. The metric derived from this methodology then quantifies the prevented expected points (PEP). 


\section{Data}

For the NFL Big Data Bowl 2024, the National Football League (NFL) supplied an extensive dataset --- spanning the first nine weeks of the 2022 season --- including game-level, play-by-play-level, player, and tackle information, and most importantly a fine-scale data set of 10 Hz tracking data within each play. The latter contains the x- and y-location, speed, acceleration, distance traveled, orientation, and direction of each of the 22 players on the field and the ball. In total, the data set comprises 12,486 plays (of varying lengths), leading to a total number of 12,187,398 tracking observations.

As our aim is the accurate prediction of the yard line at the end of any given play, a considerable amount of preprocessing is necessary to create variables that allow for said prediction. 
It is unreasonable to model the (within-play constant) EOPY as, during a play, it depends on the current position of the ball carrier. Thus, as the response variable for the subsequent analyses, we define the \textit{yards to be gained} at each frame in the play as the difference between the current x-position of the ball carrier and his x-position at the end of the play. This variable can of course easily be transformed back to a prediction for the EOPY, by adding the current ball carrier position.

As features to enable the above prediction, we start by using all raw features already contained in the tracking data, namely x- and y-coordinates, speed, acceleration, distance covered, orientation, and direction. However, before more involved feature engineering steps, we transform the coordinate system to fit the purpose of our analysis better, by i) redefining the x-variable as the x-distance to the opponent's end zone (such that all play directions are from right to left and the relevant end zone is at zero), ii) centering the y-variable such that the center of the field is at zero, and iii) modifying the direction variable, such that zero degrees represents heading straight towards the corresponding end zone.

Subsequently, we derive additional features by computing the Euclidean distance, x-distance, and y-distance to the ball carrier for all players excluding the latter. Furthermore, for defensive players only, we compute the absolute angle difference between the defender’s direction and the angle of the shortest segment between the defender and the ball carrier. Subsequently, for each frame separately, we order all players by their Euclidean distance to the ball carrier and standardize all features.


Lastly, to identify tackle events or tackle attempts jointly, i.e.\ the instant of first contact between the ball carrier and the player that attempts a tackle, either leading to a successful or failed tackle, we consider the frame for which the distance of the tackler (informed by the tackle event data set) to the ball carrier is minimal within a given play. 

As we evaluate tackles by comparing a hypothetical outcome to the true EOPY, it is unreasonable to include every play containing a tackle for the final evaluation. Specifically, we exclude all plays with penalties as this affects the final yard line of the play, which is why a comparison of the hypothetical yard line prediction to this true yard line is not reasonable. Finally, we ended up with 11,313 tackles to analyze.

\section{Methods}

In this section, we progressively present the steps required to calculate the final PEP value. We begin with estimating the full conditional density of the EOPY using a random forest, afterwards showing how we convert this into an EP value, and finally calculating the value of a real tackle by comparing its EP with the equivalent of a hypothetically missed tackle in the same situation.

\subsection{Random forest conditional density estimation}

In our initial phase of analysis, we require a robust model capable of precisely predicting the yard line at the end of a given play. Several options stand out as particularly well-suited for this task, including long-existing machine learning algorithms like random forest and XGBoost \citep{breiman2001random, XGBOOST}, and more recent artificial intelligence models such as LSTMs and transformers \citep{hochreiter1997long, vaswani2017attention}.
It is crucial that our model can effectively capture potential intricate, non-linear relationships and interactions among player positions, speeds, and other factors influencing the final EOPY. Additionally, given the abundance of features involved, we seek a model equipped with automated feature selection capabilities.

An additional requirement for the model used is the accommodation for adequate uncertainty quantification. 
This aspect is crucial to the specific task of yard-line prediction, considering the wide-ranging uncertainty that is present in football play situations. For example, consider a defender being very close to the ball carrier in the instantaneous moment before attempting a tackle: A successful tackle typically results in minimal yardage gain, while a missed tackle most definitely leads to a substantial additional gain, and potentially even to a touchdown. Accurate modeling of such a game situation demands a flexible forecasting distribution that acknowledges the multi-modal nature of outcomes. 

\begin{figure}
    \centering
    \includegraphics[width=1\textwidth]{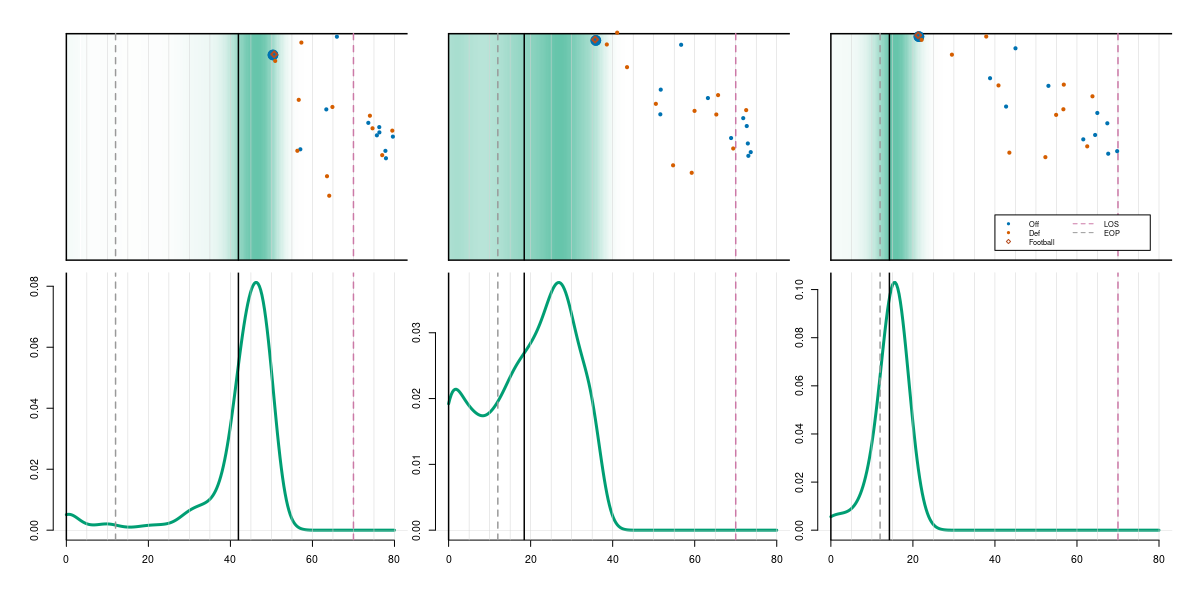}
    \caption{Three frames of an example play with conditional density estimate of the EOPY. After catching the ball, the distribution for the EOPY is concentrated as the model expects a tackle from the closest defender (left panel). The ball carrier goes on to evade a tackle leading to increased variance and bimodality in the density estimate with a lot of mass in the end zone (middle panel). Finally, at the time of tackle the distribution narrows again, as the model expects the ball carrier to make only a few more yards (right panel).}
    \label{fig:random}
\end{figure}

While frameworks like distributional trees \citep{schlosser} or NGboost \citep{duan2020ngboost} permit modeling the conditional distribution of the response --- instead of just the conditional mean --- they do so by assuming a parametric form of the response distribution. On the contrary, random forests (for regression) are defined as an ensemble of (regression) trees, where each tree is grown on a bootstrap sample of the original data set and only a random subset of features is drawn as split candidates at each split. Typically the individual tree predictions (each having a small bias but a large variance) are aggregated to obtain an unbiased prediction of the conditional mean of the response \citep{breiman2001random}. The variance in the tree prediction, stemming from bootstrap sampling of the data distribution, can however be exploited by using a non-parametric density estimation approach to obtain a distributional prediction, instead of aggregating by taking the mean. Thereby, random forests allow for fully non-parametric conditional density estimation, which is exactly what we need in the described setting. Figure~\ref{fig:random} demonstrates that the non-parametric density estimates obtained from random forests are sufficiently flexible to capture this adequately.

As we aim to value each tackle in our dataset, we have to derive a valid training procedure, such that an out-of-sample estimate of the conditional density is obtained. Therefore, we train a total of nine models, for each of which we use eight weeks of training data and the remaining week as evaluation set. In this way, in addition to obtaining out-of-sample estimates for each tackle, we are also able to evaluate model performance. The out-of-sample performance of our models is similar to existing approaches (see \citealp{Yurko}) with an average root mean squared error (RMSE) of 5.74 and a mean absolute error (MAE) of 3.13. We refrain from performing extensive hyperparameter tuning, for two reasons. First, typical procedures such as cross-validation or out-of-bag parameter tuning use metrics such as mean squared error to select the optimal parameters for a model. However, as we are interested in a full conditional density estimate, it is not clear which loss criterion should be employed for hyperparameter selection. Second, random forests are known to give excellent out-of-the-box predictions, meaning they enjoy good predictive performance with little parameter tuning (\citealp{HOML2019, curth2024random}). To this end, we decided to stick with the general default hyperparameters for random forests provided by the \texttt{R} package \texttt{ranger} \citep{ranger}.


\subsection{EP model}
As mentioned at the beginning of this section, we aim to measure the value of a tackle on an expected points scale. Therefore, it is necessary to model the expected points (EP) as a function of the EOPY. For our model, we assume, that there are seven different possible scoring outcomes. Here, we follow common practice (see \citealp{yurko2019}) and disregard that, after a touchdown, there is the possibility to gain either 0, 1, or 2 extra points, such that the events are: a touchdown (7 points), a field goal (3 points), a safety (2 points), an opponent safety (-2 points), an opponent field goal (-3 points), an opponent touchdown (-7 points), and no score (0 points). Expected points are then simply calculated as 
\begin{align*}
    \mathbb{E}[Y|X] = \sum_{y} y \cdot \mathbb{P}(Y = y|X), \quad y \in \{-7,-3,-2,0,2,3,7\}.
\end{align*}
Thus, to calculate EP we need to estimate the probabilities of each scoring outcome $Y$. We follow \citet{yurko2019} and allow these to be dependent on the specific game state $X$ at the end of a play. In contrast to \citet{yurko2019}, who use a multinomial logistic regression model for the estimation of expected points, we follow open source implementations of EP models (\citealp{fastR}) and use an XGBoost model for this multi-label classification task. A particular difficulty preventing the use of available implementations of EP models is that all features describing the game state have to be extractable from the predicted EOPYs of the random forest model. This disallows the usage of temporal information, as we are not predicting the time it takes to reach a specific EOPY. Hence, the covariates describing the game state in our EP model are the yard line of the play (adjusted LOS), yards to go, score differential, down, quarter, a home team indicator, and time outs remaining for each team. 

We train a model on play-by-play data from the 2011-2021 NFL seasons. Model estimation and tuning follow common practices such as weighting plays by score differential as well as cleverly performing cross-validation (leave-one-season-out cross-validation) in order to avoid score differential biases and account for the seasonal structure of the data (for details, we refer to \citealp{yurko2019}). The model is evaluated on data from the 2022 season and performs on par with comparable models with a mean absolute error (MAE) of 3.6391, compared to an MAE of 3.6395 from the model by \citet{fastR}.

\subsection{Tackle evaluation}

With the random forest conditional density estimation of the EOPY and the expected points model, we are able to proceed to the evaluation of tackles. From a mathematical perspective, we want to obtain the mean expected points, given the conditional distribution of the EOPY produced by our random forest. More formally, letting the mapping $g$ represent the calculation of expected points based on the EOPY $Y$ we are interested in
\begin{align}
\label{eq:avg_EP}
    \mathbb{E}(g(Y) \mid x) = \int g(y) \: \hat{f} (y \mid x) \: dy,
\end{align}
where $\hat{f} (y \mid x)$ is the estimated conditional density from the random forest. There are various ways to calculate the quantity in equation \eqref{eq:avg_EP}. From the random forest predictions of the EOPY in the first step, one could use a kernel density estimate (KDE) and proceed by evaluating the integral numerically. However, KDE relies on selecting a kernel as well as a smoothing parameter, the bandwidth. To avoid having to specify the setup of the KDE, and thereby reducing subjectivity, we take a more direct approach. We exploit the structure of the random forest and treat the individual tree predictions $\hat{y}_1, \dots, \hat{y}_{N}$ as samples from the conditional density, thus approximating the above expectation via the Monte Carlo estimate  
\begin{align}
\label{eq:MC_EP}
    \frac{1}{N} \sum_{i=1}^{N} g(\hat{y}_i).
\end{align}
Recall that, to quantify a tackle's value, we need to evaluate a hypothetical scenario. Thus, equation \eqref{eq:avg_EP} needs to be evaluated for the hypothetical scenario. We denote by $x_{removed}$ the transformed features at the time of tackle after removing the closest defender. Then, $\mathbb{E}\left[g(Y) \mid x_{removed}\right]$ allows us to analyze the yards gained by the ball carrier in the hypothetical scenario on an EP scale. 
In our framework, $\mathbb{E}\left[g(Y) \mid x_{removed}\right]$ is derived by performing two steps:
\begin{enumerate}
    \item Obtain draws $\hat y_{removed}$ of the conditional density $\hat{f} (y \mid x_{removed})$ from the random forest.
    \item Average over $g(\hat y_{removed})$ to obtain an estimate for EP (i.e.~plug in the draws $\hat y_{removed}$ into equation \eqref{eq:MC_EP}).
\end{enumerate}

\begin{figure}
    \centering
    \includegraphics[width=1\textwidth]{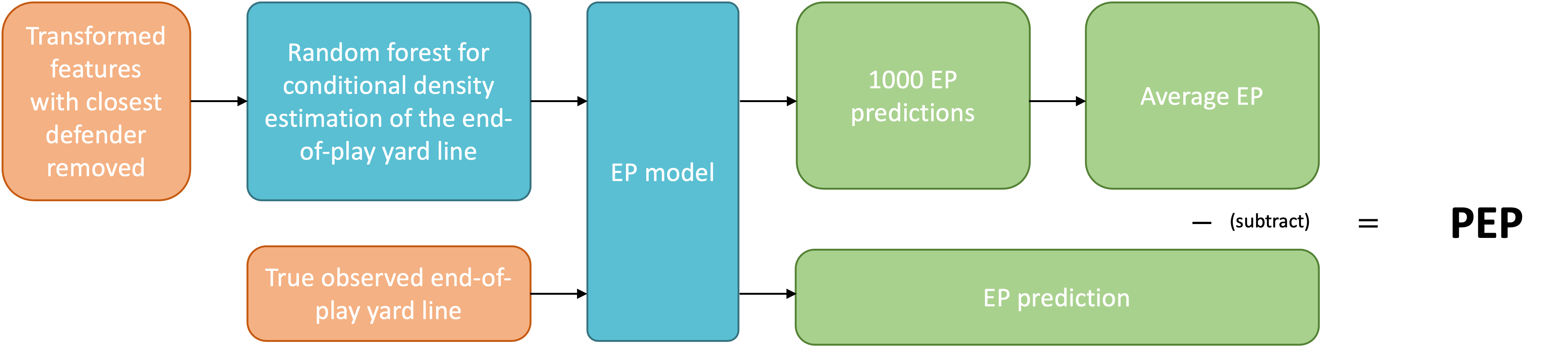}
    \caption{A graphical summary of the presented modeling pipeline.}
    \label{fig:pipeline}
\end{figure}

Since we know the true outcome, i.e.\ the true EOPY, we can compare each tackle play in our data set to the hypothetical value. Our prevented expected points metric is then computed as 
\begin{align}
\label{eq:PEP}
\text{PEP} = \mathbb{E}\left[g(Y) \mid x_{removed}\right] - g(y_0).
\end{align}
In this case, $y_0$ represents the true (observed) EOPY and thus, $g(y_0)$ provides a value for the yards gained in the real play with a tackle on an EP scale as well. Figure \ref{fig:pipeline} summarizes the necessary steps of our analysis. 

Alternatively, a value for a tackle can also be computed by comparing the predictions from the hypothetical scenario to the predictions from the actual scenario via
\begin{align}
\label{eq:PEP_alt}
\text{PEP}_{\text{alt}} = \mathbb{E}\left[g(Y) \mid x_{removed}\right] - \mathbb{E}\left[g(Y) \mid x_0\right],
\end{align}
where $x_0$ denotes the original features including the tackler at the time of tackle. Therefore, $\mathbb{E}\left[g(Y) \mid x_0\right]$ quantifies the value of the predicted yards gained in the real scenario on an EP scale. 

$\text{PEP}$ quantifies the prevented expected points by a real observed tackle which is relevant for, e.g., player evaluation as some players might over- or underperform in comparison to the model prediction. From a causal inference perspective, $\text{PEP}_{alt}$ can be regarded as a special kind of (conditional) treatment effect (\citealp{TE_review04}) representing the average expected points prevented by the tackle, given the specific game situation.
While it has its uses in analyzing the importance of tackles in given game situations, when the goal is to evaluate single players, an average effect is not desired.

\section{Results}

\subsection{PEP illustration}
\begin{figure}
    \centering
    \includegraphics[width=1\textwidth]{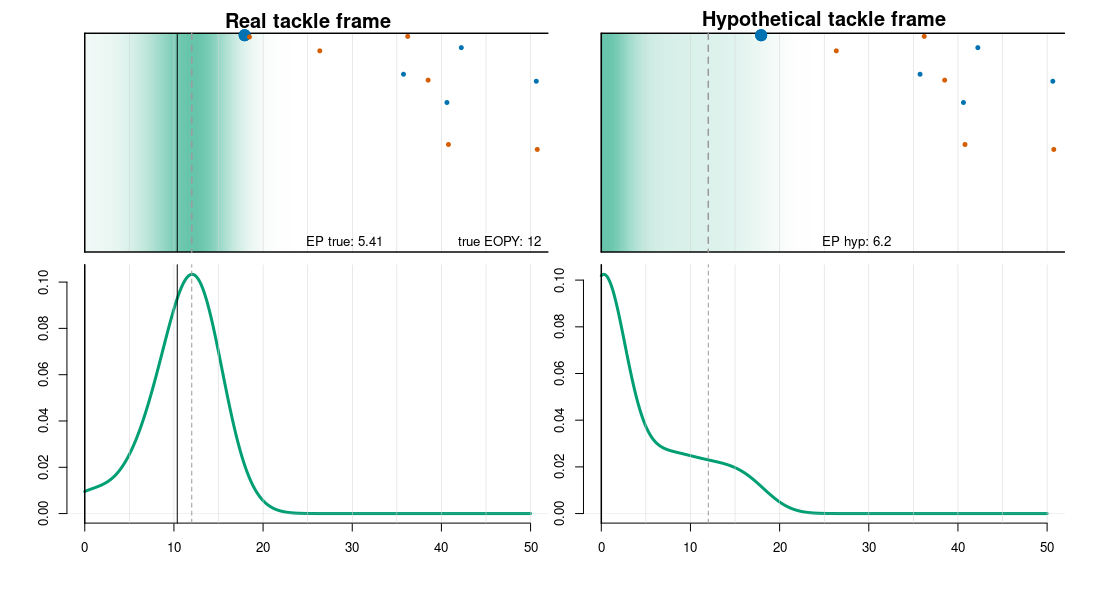}
    \caption{Left panel: Conditional density estimation for the observed frame of the tackle. Right panel: Conditional density estimate for the same frame under the hypothetical scenario of removing the tackler.}
    \label{fig:ex_play_res}
\end{figure}

We exemplify the functionality of our PEP metric using our running example play from before. The setup for all results presented in this section is based on a random forest consisting of $N = 1000$ trees. Figure~\ref{fig:ex_play_res} presents the conditional density estimation in the real and hypothetical scenarios. The gray dotted line in the figure represents the true EOPY, in this case, the 12-yard line. Given further play information (i.e.\ the initial yard line, the quarter, the down, the score differential, etc.), our trained model computes an EP value of 5.41 for this true EOPY. 
In contrast, it can be observed from the right panel of Figure~\ref{fig:ex_play_res} that 
in the hypothetical scenario, a lot of mass is in the end zone, resulting in a mean EP value of 6.2. 
To this end, the tackle in this play has a PEP (prevented expected points) value of 0.79. 

\subsection{Aggregating PEP values}

\begin{figure}
    \centering
    \includegraphics[width=0.8\textwidth]{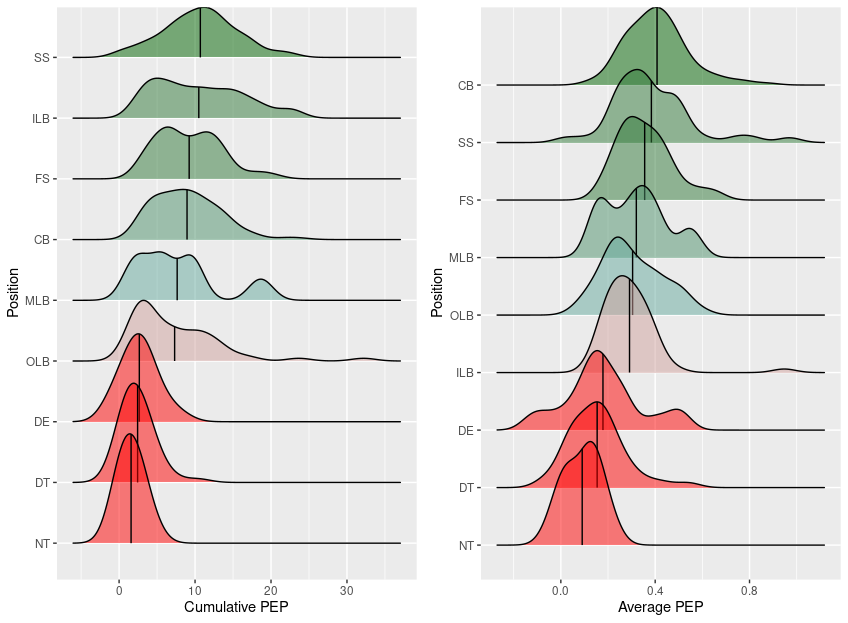}
    \caption{Densities of the cumulative (left) and the average (right) PEP values for the different position groups present in the data.}
    \label{fig:pos_groups}
\end{figure}

In the previous section, we discussed how to measure the value of a specific tackle. There are several ways to relate the value of individual tackles to the overall tackling strength of players. The most straightforward approach is to simply sum the values from all observed tackles of individual players. 
While analyzing cumulative PEP values provides interesting insights (see Appendix \ref{app:sum_pep} for more details), there are shortcomings to using it for performance analyses.
Aside from the obvious fact, that players with more tackles accumulate higher PEP values, it quickly becomes apparent that comparing players among different position groups is difficult. In particular, the left panel of Figure~\ref{fig:pos_groups} displays the densities of the sum of PEP values of players divided into the most common position groups. The figure shows that positions such as (inside) linebackers and (strong) safeties tackle the most, leading to a high cumulative PEP value. Although positions such as defensive ends (DE), nose tackles (NT), or defensive tackles (DT) are involved in almost every play, thus tackle a lot, a potential missed tackle from them can be remediated by other players (linebackers, safeties, and cornerbacks), which is why positions such as DE, NT and DT turn out to be low (sum) PEP positions. 

Another approach for analyzing a player's tackling ability is to consider the average of the PEP values. As observable from the right panel of Figure~\ref{fig:pos_groups}, the distribution of average PEP values per group differs from the cumulative PEP values. Specifically, inside linebackers (ILB) fall in the order of importance and their top spot is overtaken by defensive backs (cornerbacks (CB) and safeties (SS, FS)). This result can be explained by the fact that defensive backs are often the last players to be passed before offensive players can score a touchdown. Therefore, although these players are not involved in every play, a tackle by those players typically secures a correspondingly high amount of EP.

\subsection{Mixed effect models for PEP values}

To overcome the above-described challenges in relating tackle values to tackling strengths and appropriately attribute credit of PEP values to players, we fit a mixed effect model to these values. In particular, we try to model the PEP values for each tackle in our data set dependent on various factors and include a player's tackling ability as a random effect on the mean of the PEP values. An initial inspection of the PEP values shows that, although they seem to be roughly symmetric, they exhibit heavier tails than normally distributed random variables. Thus, in order to maintain flexibility, we use generalized additive models for location scale and shape (GAMLSS), originally developed by \citet{GAMLSS05}, for modeling PEP. That is, we allow the PEP values to come from a distribution with density $f_{PEP}$, depending on four distributional parameters $\mu,\sigma,\nu,\tau$, where $\mu$ and $\sigma$ typically represent location and scale parameters, whereas the remaining parameters characterize the shape of a distribution. In a typical GAMLSS, all four parameters can be related to explanatory variables via suitable link functions. In order to maintain interpretability, we seek to model the mean via the identity link function, allowing for random and fixed effects, i.e.
\begin{align}
\label{eq:MM_mean}
g(\mu) = \mu = \boldsymbol{X}\boldsymbol{\beta} + \boldsymbol{Z} \boldsymbol{u},
\end{align}
with design matrices $\boldsymbol{X}$ and $\boldsymbol{Z}$, fixed effects $\boldsymbol{\beta}$, and random effects $\boldsymbol{u} \sim N(0,\Sigma)$. 

\begin{figure}
    \centering
    \includegraphics[width=0.8\textwidth]{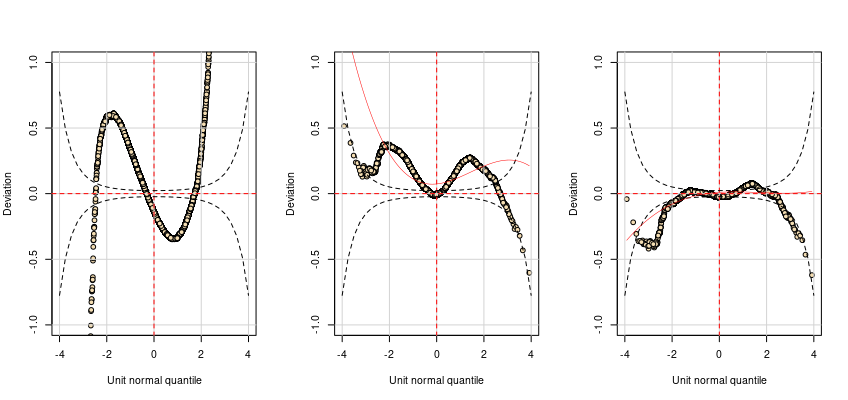}
    \caption{Wormplot for the (normalized quantile) residuals of the normal (left), the TF (middle), and the SST (right) GAMLSS model.}
    \label{fig:wormplot}
\end{figure}

Regarding our specific application, we assume random intercepts for the tackler, the ball carrier as well as the offensive team. That is, each tackler (ball carrier and offensive team) is assumed to be drawn from a normal distribution. Furthermore, we account for various attributes that influence the PEP values and estimate their respective coefficients as fixed effects. In this way, we are able to account for the positional variation we observed when aggregating or averaging PEP values. Additionally, we account for several indicators for short yardage (< 2 yards to go), fourth down, fourth quarter, and turnover as well as factors such as the pass result, and the position of the ball carrier. GAMLSS models can conveniently be fitted in \texttt{R} via the \texttt{gamlss}-package (\citealp{GAMLSS07}). To obtain a suitable model, we fit three different types of models to the data. Since initial examination suggested that PEP values exhibit heavy tails, we fit two $t$-type of families. First, a three-parameter $t$-family distribution (TF), which is symmetric around the mean $\mu$, and second, a four-parameter skew Student $t$-distribution (SST), which additionally incorporates skewness (we refer to \citealp{GAMLSS19} for details on these distributions). For model comparison, we also fit a normal distribution to the PEP values. Figure \ref{fig:wormplot} shows wormplots, i.e.\ de-trended Q-Q-plots (\citealp{wormplot01}), for the three models. Clearly, the SST model (right panel of Figure \ref{fig:wormplot}) performs best, whereas the normal distribution (left panel) is not appropriate for modeling PEP values. The final model used for PEP values can thus be written as 
\begin{equation}
\label{eq:PEP_MM}
\begin{aligned}
\text{PEP}_{i} &\sim SST(\mu_i,\sigma,\nu,\tau), &\quad i &= 1,\dots,n_{tackles}\\
\mu_i &= \boldsymbol{x}_i \boldsymbol{\beta} + T_{it} +B_{ib} + O_{io}, \\
T_t &\sim N(\mu_t,\sigma_t^2), &\quad t &= 1,\dots,n_{tacklers}, \\
B_b &\sim N(\mu_b,\sigma_b^2), &\quad b &= 1,\dots,n_{ballcarriers}, \\
O_o &\sim N(\mu_o,\sigma_o^2), &\quad o &= 1,\dots,n_{offteams}.
\end{aligned}
\end{equation}

\begin{figure}
    \centering
    \includegraphics[width=0.8\textwidth]{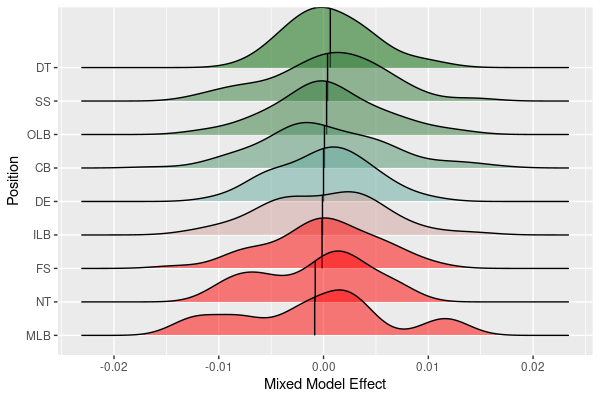}
    \caption{Densities of the player-specific random intercepts obtained from the mixed model (MM) grouped by the different positions present in the data.}
    \label{fig:pos_groups_mm}
\end{figure}

Modeling PEP values in this way has various advantages. First, Figure~\ref{fig:pos_groups_mm} shows that it is possible to eliminate positional effects on the tackle value. That is, in contrast to summing respectively averaging PEP values for players, it is possible to compare a player's tackling ability \textit{across} position groups. Second, by treating tacklers as a priori random variables, each estimate of a tackler's individual intercept is pulled towards the group mean $\mu_t$, thereby implicitly shrinking values for players with a lower number of tackles to the overall average. Third, by controlling for various play-specific attributes via fixed effects (and separate random effects), we ensure that the estimates for the tacklers accurately represent their effect on PEP values. 

Finally, to quantify uncertainty in the intercept estimates for each tackler, we use a bootstrapping approach similar to the one by \citet{nguyen}. Specifically, we cleverly resample the dataset, maintaining the intrinsic structure of the game into drives. To this end, we derive the drives of the teams in each match and instead of resampling each play individually, we resample full drives. This allows us to account for the idiosyncrasies of different drives in football. In long drives, for example, there is the need to substitute players more often for them to recover, which has to be taken into account when deciding on player usage. Then, for each of the bootstrap samples we fit the model from equation \eqref{eq:PEP_MM} in order to obtain a distribution of estimates for each tackler.

\subsection{Evaluating players}
\label{sec:player_eval}

\begin{figure}
    \centering
    \includegraphics[width=\textwidth]{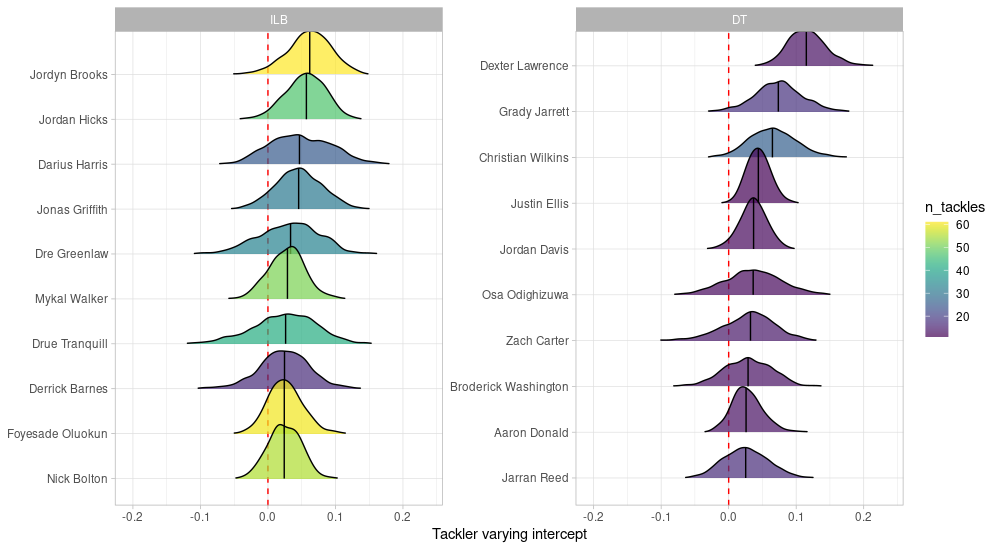}
    \caption{Distribution of the tackler effects obtained from 1000 bootstrap samples of the data. The top 10 inside linebackers (ILB, left) and defensive tackles (DT, right) are shown.}
    \label{fig:player_pos}
\end{figure}

In this section, we present the results from fitting the previously described mixed effects model to all PEP values in our data set. We analyze the results on all players that tackled more than ten times within the observed period in our data. First, in Figure \ref{fig:player_pos}, we present the bootstrap distribution of the varying intercept estimate of the top ten players from two different positions. In the left panel of Figure~\ref{fig:player_pos}, we consider inside linebackers (ILB), which traditionally tackle a lot (see also Figure~\ref{fig:pos_groups}). This is also highlighted by the colors of the densities, where lighter colors signify more tackles of the respective players. On the right-hand side, we display the top defensive tackles (DT) according to our metric. Being a low cumulative PEP position, the darker colors of the densities also indicate that these players tackle less frequently. From a domain-specific viewpoint, the results seem sensible, although they have to be taken with care as we only consider half a season's worth of tackles. Nevertheless, the top three players from the \href{https://www.nfl.com/stats/player-stats/category/tackles/2022/REG/all/defensivesolotackles/DESC}{2022 NFL tackles leaderboard} (Nick Bolton, Foyesade Oluokun, and Jordyn Brooks) are all found in top ten inside linebackers. Furthermore, these top players also exhibit less variance in their estimate for tackle value according to the densities plotted. A similar picture can be observed from the DTs, where top players such as Dexter Lawrence or Aaron Donald exhibit more narrow distributions than players such as Osa Odighizuwa or Broderick Washington who surprisingly appear in the top 10 of our ranking. 

Finally, we present a ranking of the top 20 tacklers independent of their position in Table \ref{table:results}. Specifically, the players are ranked based on the median of the varying intercept from the bootstrap density estimation of the mixed effects model. We observe a rather diverse set of players concerning their positions. The table contains players from each of the three sets of position groups that build the defense in American Football. We observe defensive liners (DT, DE), linebackers (ILB, OLB) as well as defensive backs (CB, SS, FS). In that sense, the ranking differs from one based on cumulative PEP values (compare to Table \ref{table:sum_pep_res} in Appendix \ref{app:sum_pep}). Interestingly, the top 10 are populated mostly with cornerbacks (CB), which typically is a position that is highly involved in passing plays. Usually, the task of a cornerback is to prevent opponents from catching the ball. However, when they fail to do so, they often find opponents in positions where tackles are urgently needed. In Appendix \ref{app:rush_vs_pass}, we discuss this problem in more detail and provide additional insights into this topic.

\includepdf[scale=0.85,addtolist={1,table,Results table,table:results}]{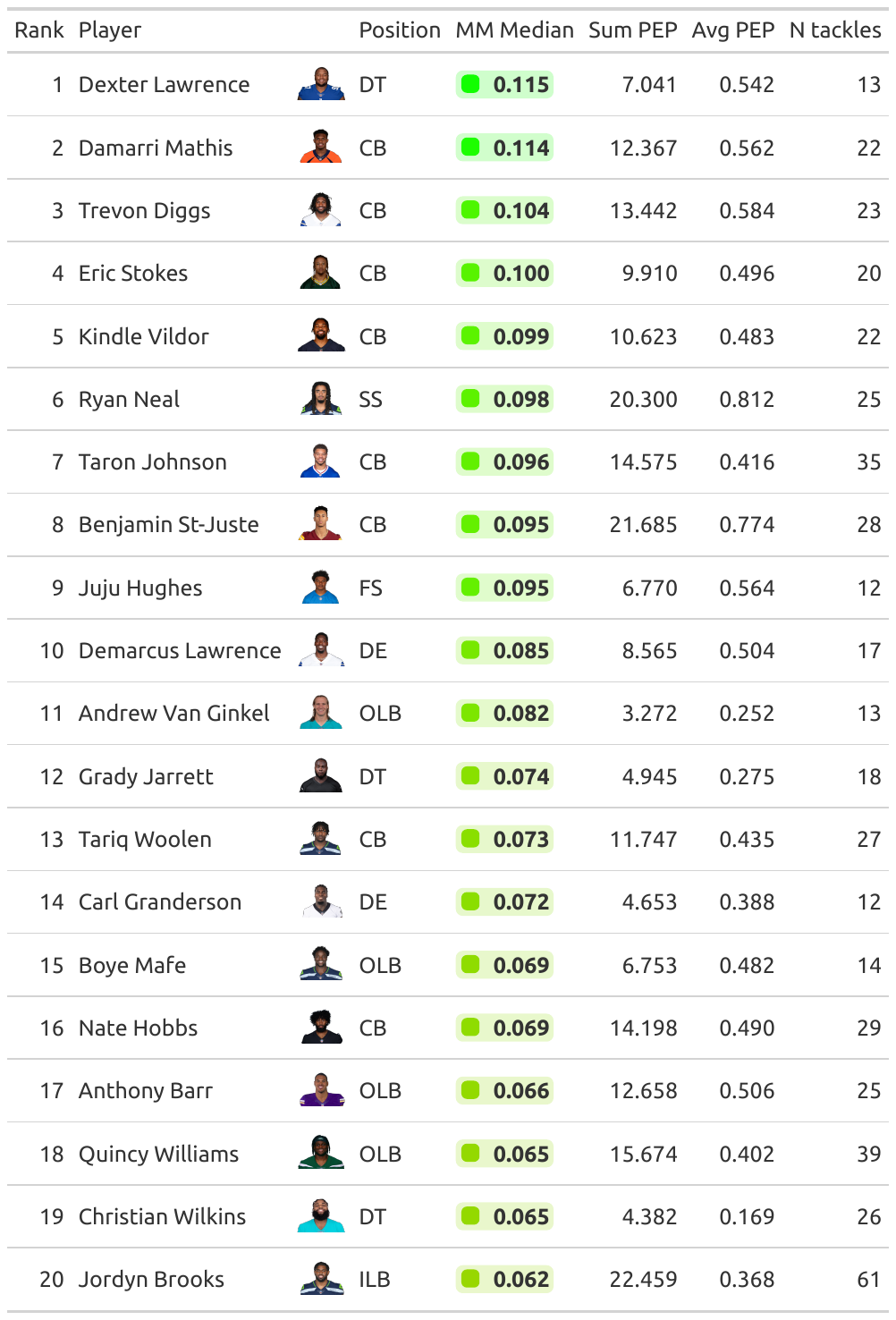}

\section{Discussion}
\label{sec:discussion}

In this contribution, we developed the metric PEP for quantifying the value of tackles, thus going beyond simple summary statistics by directly assessing the impact of tackles in a given game situation. The metric allows practitioners to evaluate players, particularly in terms of their tackling abilities. Our approach uses a within-play conditional density estimation of the EOPY, serving as a basis for the evaluation of tackle performances measured by prevented expected points. 
Importantly, our method incorporates distributional information, i.e.\ heteroscedasticity and multimodality, which would be lost when solely relying on point predictions. Therefore, the uncertainty can propagate to the level of expected points, leading to an accurate quantification of expected points prevented by the tackle.



A drawback of our current approach is that missed tackles of players are not punished by our metric as we only depict real tackles (apart from the fact that players that often miss tackles will not have a high number of tackles, thus it is hard for them to accumulate high PEP values). However, our approach is potentially also extendable to missed tackles. In that context, we would need to quantify the EOPY if the tackles were not missed but made. Comparing that value with the real EOPY would give a metric quantifying missed tackles as well. In Appendix \ref{app:missed}, we outline how to account for missed tackles and provide a short analysis thereof. 
Another caveat relates to the type of play. 
While run plays do not present any issues regarding interpretability, the situation is somewhat different for passing plays. Here, it is possible that, for example, cornerbacks generate positive PEP values simply by allowing catches by receivers and then tackling them afterward. This could result in a bias towards cornerbacks who allow catches, which is not the intended purpose of the PEP value. Therefore, special caution should be exercised when evaluating the position groups of defensive backs. Additionally, a separate analysis of run and pass plays could be beneficial in the future when a larger data set is available. We tackle this problem in more detail in Appendix \ref{app:rush_vs_pass}.

While we have focused on the defense, PEP values could also be used to assess offensive player performances. In particular, our models could identify ball carriers, most often running backs, who do not lose many EPs by being tackled, i.e.\ who only get tackled in situations in which their hypothetical EOPY is similar to the real EOPY where they got tackled.

In summary, the metric we have developed can serve as an additional piece of the puzzle in the overall evaluation of (defensive) players and may gain practical relevance in the process of scouting players and opponents.

\section{Acknowledgements}
We would like to thank the organizers of the NFL Big Data Bowl 2024 for setting up this competition and providing access to the data.

\bibliographystyle{apalike}
\bibliography{references}

\begin{thebibliography}{}

\bibitem[Adam et~al., 2024]{adam2024markov}
Adam, T., {\"O}tting, M., and Michels, R. (2024).
\newblock Markov-switching decision trees.
\newblock {\em AStA Advances in Statistical Analysis}, pages 1--16.

\bibitem[Boehmke and Greenwell, 2019]{HOML2019}
Boehmke, B. and Greenwell, B. (2019).
\newblock {\em Hands-On Machine Learning with R}.
\newblock Chapman \& Hall/CRC The R Series. CRC Press.

\bibitem[Breiman, 2001]{breiman2001random}
Breiman, L. (2001).
\newblock Random forests.
\newblock {\em Machine learning}, 45:5--32.

\bibitem[Brill et~al., 2024]{brill2023analytics}
Brill, R.~S., Yurko, R., and Wyner, A.~J. (2024).
\newblock Analytics, have some humility: a statistical view of fourth-down decision making.
\newblock arXiv:2311.03490.

\bibitem[Buuren and Fredriks, 2001]{wormplot01}
Buuren, S.~v. and Fredriks, M. (2001).
\newblock Worm plot: a simple diagnostic device for modelling growth reference curves.
\newblock {\em Statistics in Medicine}, 20(8):1259--1277.

\bibitem[Carl and Baldwin, 2023]{fastR}
Carl, S. and Baldwin, B. (2023).
\newblock {\em nflfast{R}: Functions to Efficiently Access {NF}L Play by Play Data}.
\newblock \url{https://github.com/nflverse/nflfastR}.

\bibitem[Chen and Guestrin, 2016]{XGBOOST}
Chen, T. and Guestrin, C. (2016).
\newblock {XGBoost: A Scalable Tree Boosting System}.
\newblock In {\em Proceedings of the 22nd {ACM} {SIGKDD} International Conference on Knowledge Discovery and Data Mining}. {ACM}.

\bibitem[Chu et~al., 2020]{ChuReyersThomsonWu+2020+121+132}
Chu, D., Reyers, M., Thomson, J., and Wu, L.~Y. (2020).
\newblock Route identification in the {N}ational {F}ootball {L}eague.
\newblock {\em Journal of Quantitative Analysis in Sports}, 16(2):121--132.

\bibitem[Curth et~al., 2024]{curth2024random}
Curth, A., Jeffares, A., and van~der Schaar, M. (2024).
\newblock Why do random forests work? {U}nderstanding tree ensembles as self-regularizing adaptive smoothers.
\newblock {\em arXiv:2402.01502}.

\bibitem[Deshpande and Evans, 2020]{deshpande2020expected}
Deshpande, S.~K. and Evans, K. (2020).
\newblock Expected hypothetical completion probability.
\newblock {\em Journal of Quantitative Analysis in Sports}, 16(2):85--94.

\bibitem[DFL, 2024]{dfl}
DFL (2024).
\newblock The source of all – the official match data.
\newblock https://www.dfl.de/en/topics/match-data/official-match-data/.

\bibitem[Duan et~al., 2020]{duan2020ngboost}
Duan, T., Anand, A., Ding, D.~Y., Thai, K.~K., Basu, S., Ng, A., and Schuler, A. (2020).
\newblock Ngboost: Natural gradient boosting for probabilistic prediction.
\newblock In {\em International conference on machine learning}, pages 2690--2700. PMLR.

\bibitem[Dutta et~al., 2020]{DuttaYurkoVentura+2020+143+161}
Dutta, R., Yurko, R., and Ventura, S.~L. (2020).
\newblock Unsupervised methods for identifying pass coverage among defensive backs with {NFL} player tracking data.
\newblock {\em Journal of Quantitative Analysis in Sports}, 16(2):143--161.

\bibitem[Eager and Seth, 2023]{eager2023investigating}
Eager, E. and Seth, T. (2023).
\newblock Investigating trade-offs made by {A}merican football linebackers using tracking data.
\newblock {\em Journal of Quantitative Analysis in Sports}, 19(3):171--185.

\bibitem[Fernandes et~al., 2020]{joash2020predicting}
Fernandes, C.~J., Yakubov, R., Li, Y., Prasad, A.~K., and Chan, T.~C. (2020).
\newblock Predicting plays in the {N}ational {F}ootball {L}eague.
\newblock {\em Journal of Sports Analytics}, 6(1):35--43.

\bibitem[Forcher et~al., 2022]{forcher2022use}
Forcher, L., Altmann, S., Forcher, L., Jekauc, D., and Kempe, M. (2022).
\newblock The use of player tracking data to analyze defensive play in professional soccer-a scoping review.
\newblock {\em International Journal of Sports Science \& Coaching}, 17(6):1567--1592.

\bibitem[Goes et~al., 2021]{goes2021tactics}
Goes, F.~R., Brink, M.~S., Elferink-Gemser, M.~T., Kempe, M., and Lemmink, K.~A. (2021).
\newblock The tactics of successful attacks in professional association football: large-scale spatiotemporal analysis of dynamic subgroups using position tracking data.
\newblock {\em Journal of Sports Sciences}, 39(5):523--532.

\bibitem[Heiny and Blevins, 2011]{heiny2011predicting}
Heiny, E.~L. and Blevins, D. (2011).
\newblock Predicting the {A}tlanta {F}alcons play-calling using discriminant analysis.
\newblock {\em Journal of Quantitative Analysis in Sports}, 7(3):Article 2.

\bibitem[Hochreiter and Schmidhuber, 1997]{hochreiter1997long}
Hochreiter, S. and Schmidhuber, J. (1997).
\newblock Long short-term memory.
\newblock {\em Neural computation}, 9(8):1735--1780.

\bibitem[Imbens, 2004]{TE_review04}
Imbens, G.~W. (2004).
\newblock Nonparametric estimation of average treatment effects under exogeneity: A review.
\newblock {\em The Review of Economics and Statistics}, 86(1):4--29.

\bibitem[Kovalchik, 2023]{kovalchik2023player}
Kovalchik, S.~A. (2023).
\newblock Player tracking data in sports.
\newblock {\em Annual Review of Statistics and Its Application}, 10:677--697.

\bibitem[Michels and Langrock, 2023]{michels2023nonparametric}
Michels, R. and Langrock, R. (2023).
\newblock Nonparametric estimation of multivariate hidden markov models using tensor-product {B}-splines.
\newblock {\em arXiv preprint arXiv:2302.06510}.

\bibitem[M{\"u}ller et~al., 2021]{muller2021pivot}
M{\"u}ller, O., Caron, M., D{\"o}ring, M., Heuwinkel, T., and Baumeister, J. (2021).
\newblock {PIVOT}: a parsimonious end-to-end learning framework for valuing player actions in handball using tracking data.
\newblock In {\em International Workshop on Machine Learning and Data Mining for Sports Analytics}, pages 116--128. Springer.

\bibitem[Nguyen et~al., 2024]{nguyen2024fractional}
Nguyen, Q., Jiang, R., Ellingwood, M., and Yurko, R. (2024).
\newblock Fractional tackles: Leveraging player tracking data for within-play tackling evaluation in {A}merican football.
\newblock {\em arXiv preprint arXiv:2403.14769}.

\bibitem[Nguyen et~al., 2023]{nguyen}
Nguyen, Q., Yurko, R., and Matthews, G.~J. (2023).
\newblock Here comes the {STRAIN}: analyzing defensive pass rush in {A}merican football with player tracking data.
\newblock {\em The American Statistician}.
\newblock In press.

\bibitem[{\"O}tting and Karlis, 2023]{otting2023football}
{\"O}tting, M. and Karlis, D. (2023).
\newblock Football tracking data: a copula-based hidden {M}arkov model for classification of tactics in football.
\newblock {\em Annals of Operations Research}, 325(1):167--183.

\bibitem[Reyers and Swartz, 2023]{reyers2021quarterback}
Reyers, M. and Swartz, T.~B. (2023).
\newblock Quarterback evaluation in the {N}ational {F}ootball {L}eague using tracking data.
\newblock {\em AStA Advances in Statistical Analysis}, 107(1-2):327--342.

\bibitem[Rigby and Stasinopoulos, 2005]{GAMLSS05}
Rigby, R. and Stasinopoulos, M. (2005).
\newblock Generalized additive models for location, scale and shape,(with discussion).
\newblock {\em Applied Statistics}, 54:507--554.

\bibitem[Rigby et~al., 2019]{GAMLSS19}
Rigby, R., Stasinopoulos, M., Heller, G., and De~Bastiani, F. (2019).
\newblock {\em Distributions for Modeling Location, Scale, and Shape: Using GAMLSS in R}.
\newblock Chapman \& Hall/CRC The R Series. CRC Press.

\bibitem[Schlosser et~al., 2018]{schlosser}
Schlosser, L., Hothorn, T., Stauffer, R., and Zeileis, A. (2018).
\newblock Distributional regression forests for probabilistic precipitation forecasting in complex terrain.
\newblock {\em The Annals of Applied Statistics}, 13.

\bibitem[Stasinopoulos and Rigby, 2007]{GAMLSS07}
Stasinopoulos, M. and Rigby, R. (2007).
\newblock Generalized additive models for location scale and shape (gamlss) in {R}.
\newblock {\em Journal of Statistical Software}, 23(7):1–46.

\bibitem[Van~Haaren, 2021]{vanhaaren}
Van~Haaren, J. (2021).
\newblock Industry-leaders in football’s use of data intelligence.
\newblock \url{https://www.scisports.com/state-of-the-football-analytics-industry-in-2021/}.

\bibitem[Vaswani et~al., 2017]{vaswani2017attention}
Vaswani, A., Shazeer, N., Parmar, N., Uszkoreit, J., Jones, L., Gomez, A.~N., Kaiser, {\L}., and Polosukhin, I. (2017).
\newblock Attention is all you need.
\newblock {\em Advances in neural information processing systems}, 30.

\bibitem[Wright and Ziegler, 2017]{ranger}
Wright, M.~N. and Ziegler, A. (2017).
\newblock {ranger}: A fast implementation of random forests for high dimensional data in {C++} and {R}.
\newblock {\em Journal of Statistical Software}, 77(1):1--17.

\bibitem[Yam and Lopez, 2019]{yam2019lost}
Yam, D.~R. and Lopez, M.~J. (2019).
\newblock What was lost? {A} causal estimate of fourth down behavior in the {N}ational {F}ootball {L}eague.
\newblock {\em Journal of Sports Analytics}, 5(3):153--167.

\bibitem[Yurko et~al., 2020]{Yurko}
Yurko, R., Matano, F., Richardson, L.~F., Granered, N., Pospisil, T., Pelechrinis, K., and Ventura, S.~L. (2020).
\newblock Going deep: models for continuous-time within-play valuation of game outcomes in {A}merican football with tracking data.
\newblock {\em Journal of Quantitative Analysis in Sports}, 16(2):163--182.

\bibitem[Yurko et~al., 2024]{yurko2024nfl}
Yurko, R., Nguyen, Q., and Pelechrinis, K. (2024).
\newblock Nfl ghosts: A framework for evaluating defender positioning with conditional density estimation.
\newblock {\em arXiv preprint arXiv:2406.17220}.

\bibitem[Yurko et~al., 2019]{yurko2019}
Yurko, R., Ventura, S., and Horowitz, M. (2019).
\newblock nfl{WAR}: a reproducible method for offensive player evaluation in football.
\newblock {\em Journal of Quantitative Analysis in Sports}, 15(3):163--183.

\end{thebibliography}

\newpage
\appendix

\section{Further Results}

\subsection{Cumulative PEP values}
\label{app:sum_pep}

Table \ref{table:sum_pep_res} displays the top 20 players based on their cumulative PEP values. Intuitively, the results from the PEP values seem reasonable. A simple sanity check is to compare these results with conventional tackle rankings as e.g.~provided by the NFL via their \href{https://www.nfl.com/stats/player-stats/category/tackles/2022/REG/all/defensivecombinetackles/DESC}{2022 NFL tackles leaderboard}. Six out of the top ten players (based on combined tackles) of this leaderboard are also found in our top 20 ranking. While this result is to some extent reassuring, it further indicates the shortcomings of using cumulative PEP values as indicators for tackle value. In the top 20, we find mostly linebackers and safeties, whereas not a single defensive liner is present. Thus, this metric fails to account for the ability of defensive linemen to stop forward movement in critical situations consistently. 

\begin{table}[!htbp] \centering 
  \caption{Top 20 players with respect to cumulative PEP values.} 
  \label{table:sum_pep_res} 
\begin{tabular}{@{\extracolsep{5pt}} cllrr} 
\\[-1.8ex]\hline 
\hline \\[-1.8ex] 
Rank & Player & Position & Sum PEP & N tackles \\ 
\hline \\[-1.8ex] 
1 & Pete Werner & OLB & 32.190 & 63 \\ 
2 & Coby Bryant & CB & 23.659 & 27 \\ 
3 & Zaire Franklin & OLB & 23.649 & 62 \\ 
4 & Foyesade Oluokun & ILB & 23.020 & 60 \\ 
5 & Jordan Hicks & ILB & 22.776 & 47 \\ 
6 & Jordyn Brooks & ILB & 22.459 & 61 \\ 
7 & Kerby Joseph & SS & 22.283 & 23 \\ 
8 & Benjamin St-Juste & CB & 21.685 & 28 \\ 
9 & Bobby Okereke & ILB & 20.369 & 63 \\ 
10 & Ryan Neal & SS & 20.300 & 25 \\ 
11 & Adrian Amos & FS & 20.179 & 31 \\ 
12 & Tremaine Edmunds & ILB & 19.035 & 47 \\ 
13 & Cody Barton & MLB & 18.653 & 45 \\ 
14 & Derwin James & FS & 18.625 & 55 \\ 
15 & Devin Lloyd & ILB & 18.503 & 54 \\ 
16 & Rashaan Evans & ILB & 18.168 & 47 \\ 
17 & D.J. Reed & CB & 18.043 & 36 \\ 
18 & Chuck Clark & SS & 17.430 & 43 \\ 
19 & Budda Baker & SS & 17.398 & 49 \\ 
20 & C.J. Mosley & ILB & 17.390 & 56 \\ 
\hline \\[-1.8ex] 
\end{tabular} 
\end{table} 


\subsection{Mixed effects model results for all positions}

\begin{figure}
    \centering
    \includegraphics[width=1\textwidth]{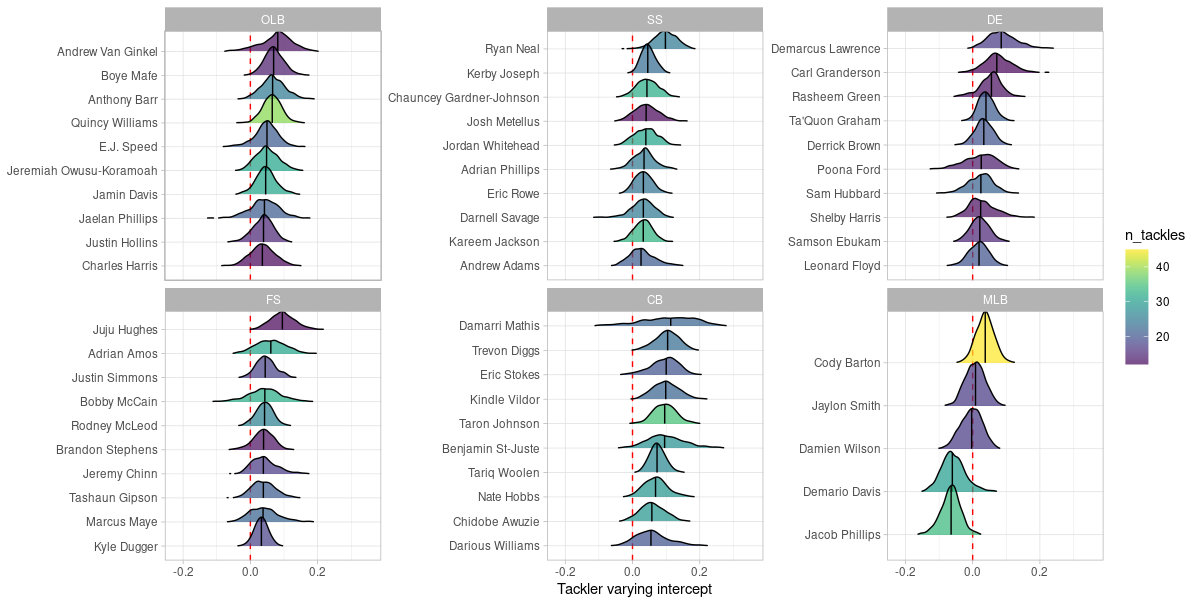}
    \caption{Distribution of the top 10 (if available) tackler effects for the remaining position groups.}
    \label{fig:player_pos_rest}
\end{figure}

In addition to the results presented in section \ref{sec:player_eval}, we provide further results on the effect distribution for players in other position groups. Figure \ref{fig:player_pos_rest} is in the pendant to Figure \ref{fig:player_pos} and displays the distribution of the mixed effect model estimates for the top ten players in the remaining position groups. Note that we do not observe more than five nominal middle linebackers (MLB) with more than ten tackles, hence only five players are shown. In principle, the results are similar to the observations of \ref{sec:player_eval}.

\subsection{Rushing vs. passing plays}
\label{app:rush_vs_pass}

As pointed out in the discussion (Section \ref{sec:discussion}), the type of play (i.e.~pass vs.~run) affects our final estimate of player strength. While it would be possible to account for the play type in our mixed model specification, it is not (at least not directly) possible to characterize a cornerback's ability. That is, we have no way of determining whether the cornerback allowed a catch that he should have already stopped earlier. Thus it is questionable, whether passing plays should be taken into account when analyzing PEP values. In this section, we briefly address this issue. To this end, we filtered tackles resulting from run plays as identified by the play type variable from play-by-play data (leaving us with 5889 tackles to analyze) and refit the model. 

Table \ref{table:run_plays} presents the result of this analysis. Interestingly, a cornerback pops up on top of our table. However, in comparison to Table \ref{table:results}, we observe fewer cornerbacks in the top spots. We again stress that looking only at run plays reduces the number of tackles in our dataset and therefore also the number of tackles of each individual. In order to be consistent with the previous results, we displayed only players, who were able to tackle more than ten times within run plays. Doing so excludes, for example, Dexter Lawrence (the top player in our full dataset, see table \ref{table:results}), for whom we observed exactly ten run-play-tackles.  


\begin{table}[!htbp] \centering 
  \caption{Top 20 players considering only run plays.} 
  \label{table:run_plays} 
\begin{tabular}{@{\extracolsep{5pt}} cllrrrr} 
\\[-1.8ex]\hline 
\hline \\[-1.8ex] 
Rank & Player & Position & MM Median & Sum PEP & Avg PEP & N tackles \\ 
\hline \\[-1.8ex] 
1 & Jeff Okudah & CB & 0.156 & 8.801 & 0.587 & 15 \\ 
2 & Demarcus Lawrence & DE & 0.138 & 6.707 & 0.479 & 14 \\ 
3 & Grady Jarrett & DT & 0.111 & 2.226 & 0.159 & 14 \\ 
4 & Kareem Jackson & SS & 0.095 & 4.624 & 0.289 & 16 \\ 
5 & Alex Anzalone & ILB & 0.093 & 5.931 & 0.282 & 21 \\ 
6 & Shaquil Barrett & OLB & 0.089 & 4.864 & 0.324 & 15 \\ 
7 & Brandon Jones & SS & 0.086 & 3.059 & 0.278 & 11 \\ 
8 & Kenny Moore & CB & 0.080 & 4.123 & 0.317 & 13 \\ 
9 & Jamin Davis & OLB & 0.077 & 3.115 & 0.195 & 16 \\ 
10 & Denzel Perryman & ILB & 0.075 & 3.895 & 0.216 & 18 \\ 
11 & Fred Warner & ILB & 0.075 & 3.217 & 0.214 & 15 \\ 
12 & Jordyn Brooks & ILB & 0.073 & 6.692 & 0.231 & 29 \\ 
13 & Christian Wilkins & DT & 0.072 & 2.700 & 0.117 & 23 \\ 
14 & Roquan Smith & ILB & 0.071 & 9.268 & 0.211 & 44 \\ 
15 & Grant Delpit & SS & 0.061 & 5.055 & 0.281 & 18 \\ 
16 & Osa Odighizuwa & DT & 0.060 & 3.425 & 0.285 & 12 \\ 
17 & E.J. Speed & OLB & 0.060 & 10.183 & 0.536 & 19 \\ 
18 & Jordan Hicks & ILB & 0.060 & 4.696 & 0.224 & 21 \\ 
19 & Marcus Maye & FS & 0.060 & 4.806 & 0.370 & 13 \\ 
20 & Vonn Bell & SS & 0.059 & 3.737 & 0.340 & 11 \\ 
\hline \\[-1.8ex] 
\end{tabular} 
\end{table} 


\subsection{Adding missed tackles}
\label{app:missed}

\begin{figure}
    \centering
    \includegraphics[width=\textwidth]{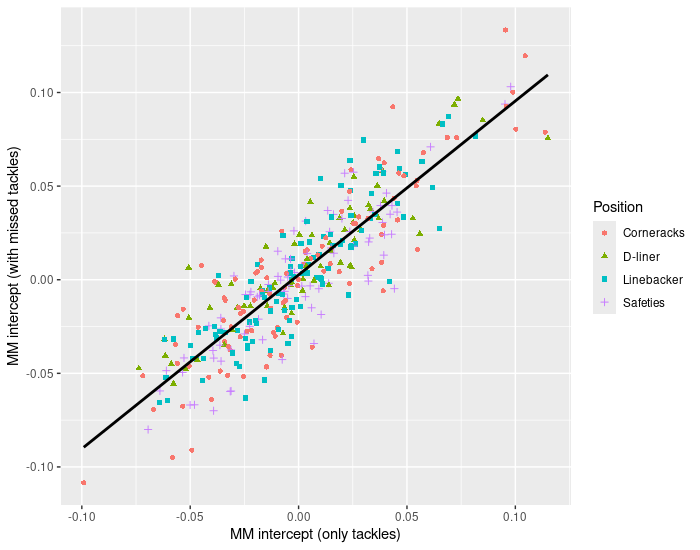}
    \caption{Relationship between mixed model tackler effect estimates with and without missed tackles. A strong linear correlation (r = 0.88137) is observable.}
    \label{fig:missed_vs_real}
\end{figure}

\begin{figure}
    \centering
    \includegraphics[width=\textwidth]{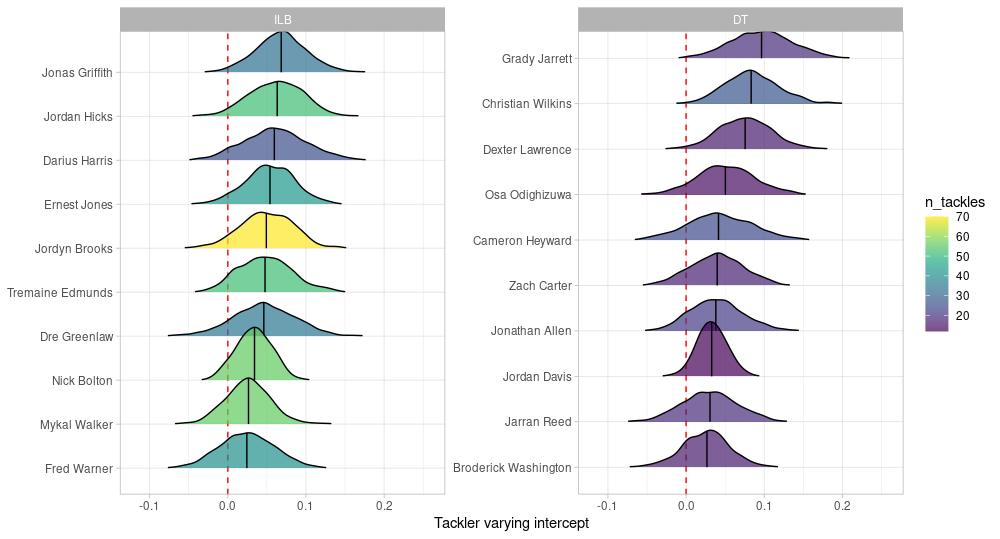}
    \caption{Distribution of the top 10 inside linebackers (ILB, left) and defensive tackles (DT, right) when adding missed tackles. Results are similar to results from Figure \ref{fig:player_pos}.}
    \label{fig:player_pos_wm}
\end{figure}

Quantifying the value of missed tackles is an important aspect when analyzing a player's tackling ability. As mentioned in the discussion, it is possible to extend our framework to analyzing missed tackles. To this end, we could treat missed tackles as tackles, predict the EOPY, and obtain a value for this hypothetical tackle on the EP scale. This value could again be compared to the real outcome allowing us to derive a missed tackle PEP value. However, this relies on accurately identifying tackle opportunities respectively missed tackles, which is not an easy task. The big data bowl provides information on missed tackles --- these have been obtained from the data provider PFF --- within the timeframe of our data. Compared to observed tackles (11,313), the number of missed tackles in the data is substantially lower (1669). Thus, we believe that solely analyzing missed tackles with this small data set is inappropriate. However, we can combine the PEP values from missed tackles and real tackles, refit the mixed effects model for the PEP values, and analyze the varying intercepts for the tacklers. In general, the results from adding missed tackles are similar to the ones obtained without them. Figures~\ref{fig:missed_vs_real} and \ref{fig:player_pos_wm} provide a visual confirmation of that. However, since identifying missed tackles is intricate, it is unclear whether the missed tackles distributions with respect to players and positions in our data are accurate and reflect the true missed tackles events distribution. Therefore, we refrain from adding them to the main analysis in this work.

\end{document}